\documentclass[journal]{vgtc}                % final (journal style)
\ifpdf%                                % if we use pdflatex
  \pdfoutput=1\relax                   % create PDFs from pdfLaTeX
  \usepackage{graphicx}                % allow us to embed graphics files
  \DeclareGraphicsExtensions{.pdf,.png,.jpg,.jpeg} % for pdflatex we expect .pdf, .png, or .jpg files
\else%                                 % else we use pure latex
  \usepackage{graphicx}                % allow us to embed graphics files
  \DeclareGraphicsExtensions{.eps}     % for pure latex we expect eps files
\fi%

\graphicspath{{figures/}{pictures/}{images/}{./}} % where to search for the images
\usepackage[dvipsnames, svgnames, x11names]{xcolor} 
\usepackage{microtype}                 % use micro-typography (slightly more compact, better to read)
\PassOptionsToPackage{warn}{textcomp}  % to address font issues with \textrightarrow
\usepackage{textcomp}                  % use better special symbols
\usepackage{mathptmx}                  % use matching math font
\usepackage{times}                     % we use Times as the main font
\usepackage{cite}                      % needed to automatically sort the references
\usepackage{tabu}                      % only used for the table example
\usepackage{booktabs} 
\usepackage{amsmath}
\usepackage{multirow} 

\usepackage{hyperref}  %用于正文引用更加智能一点，自动告诉我们引用的内容是图，表还是公式
\hypersetup{pdfborder={0 0 0} }  %we don't want those silly boxes for links

\usepackage{enumitem}
\usepackage{amsmath}
\usepackage{setspace}

\usepackage{color}
\usepackage{colortbl}
\usepackage{makecell}%导入宏包
\usepackage{array,hhline}

\usepackage[marginal]{footmisc}

\usepackage{soul}
\sethlcolor{YellowOrange}

\soulregister\cite7 % 针对\cite命令
\soulregister\autoref7 % 针对\citep命令
\soulregister\ref7 % 针对\ref命令
\onlineid{1082}

\vgtccategory{Research}

\vgtcpapertype{Algorithm}

\title{Preserving Minority Structures in Graph Sampling}

\author{Ying Zhao, Haojin Jiang, Qi'an Chen, Yaqi Qin, Huixuan Xie, Yitao Wu\\ Shixia Liu, Zhiguang Zhou, Jiazhi Xia and Fangfang Zhou}

\authorfooter{
\item
 Ying Zhao, Haojin Jiang, Qi'an Chen, Yaqi Qin, Huixuan Xie, Yitao Wu, Jiazhi Xia, and Fangfang Zhou are with School of Computer Sciences and Engineering, Central South University, China.
 E-mail: $\left \{ zhaoying,gallows,204712126,0921160110,204711057 \right \}$@csu.edu.cn, $\left \{ wuyitao51,xiajiazhi,zff \right \}$@csu.edu.cn.
 \item
 Shixia Liu is with School of Software, Tsinghua University, China.
 E-mail: shixia@tsinghua.edu.cn
 \item 
 Zhiguang Zhou is with School of Information, Zhejiang University of Finance and Economics, China. 
 E-mail: zhgzhou1983@zufe.edu.cn.
 \item
 Jiazhi Xia and Fangfang Zhou are corresponding authors.

}

\abstract{Sampling is a widely used graph reduction technique to accelerate graph computations and simplify graph visualizations. By comprehensively analyzing the literature on graph sampling, we assume that existing algorithms cannot effectively preserve minority structures that are rare and small in a graph but are very important in graph analysis. In this work, we initially conduct a pilot user study to investigate representative minority structures that are most appealing to human viewers. We then perform an experimental study to evaluate the performance of existing graph sampling algorithms regarding minority structure preservation. Results confirm our assumption and suggest key points for designing a new graph sampling approach named mino-centric graph sampling (MCGS). In this approach, a triangle-based algorithm and a cut-point-based algorithm are proposed to efficiently identify minority structures. A set of importance assessment criteria are designed to guide the preservation of important minority structures. Three optimization objectives are introduced into a greedy strategy to balance the preservation between minority and majority structures and suppress the generation of new minority structures. A series of experiments and case studies are conducted to evaluate the effectiveness of the proposed MCGS.
} % end of abstract

\keywords{Graph sampling, graph visualization, node-link diagram}

\begin{document}
%% The ``\maketitle'' command must be the first command after the
%% ``\begin{document}'' command. It prepares and prints the title block.
%% the only exception to this rule is the \firstsection command
\firstsection{Introduction}
\maketitle
%% \section{Introduction} %for journal use above \firstsection{..} instead
Graphs contain plentiful structures that can be categorized from diverse perspectives, such as normal or abnormal \cite{B10} and central or periphery \cite{A22}. In this work, we categorize graph structures into majority and minority depending on their occurrence frequencies and sizes in a graph. Majority structures refer to those frequently occurring (e.g., frequent subgraphs) or large (e.g., communities). Minority structures are those rarely occurring and containing only a few nodes (e.g., extremely high degree nodes and bridges between communities). Both categories are important in graph analysis and are of great concern in various fields, such as community detection \cite{A20}, frequent subgraph mining \cite{B43}, spammers identification \cite{B10}, and bridge vulnerability estimation \cite{A21}.

\par{Sampling is an efficient graph reduction technique \cite{A4,A5,addEOD,addVisual}. Many graph sampling algorithms have been proposed to reduce graph sizes while preserving structures \cite{A6,A7,A8}. They are particularly useful in accelerating graph computations \cite{A9} and simplifying graph visualizations \cite{A10}. However, we assume that the existing algorithms tend to preserve majority structures but overlook minority structures, because the influence of majority structures on the representativeness of the original graph is considerable for measurable metrics and visual perception, whereas that of minority ones is negligible. For example, the algorithms tend to select the nodes with common degrees far more than the nodes with rare degrees to maintain the power law of degree distribution \cite{A11}. Human viewers prefer to observe large structures in advance but may ignore small structures when judging whether a sample is visually similar to the original graph \cite{A12,A13}.}

\par
\begin{figure*}[t]
	\centering
	%\vspace{-0.6cm}  %调整图片与上文的垂直距离	
	\setlength{\abovecaptionskip}{1pt}   %调整图片标题与图距离	
	\setlength{\belowcaptionskip}{-0.6cm}   %调整图片标题与下文距离
	\includegraphics[width=\linewidth]{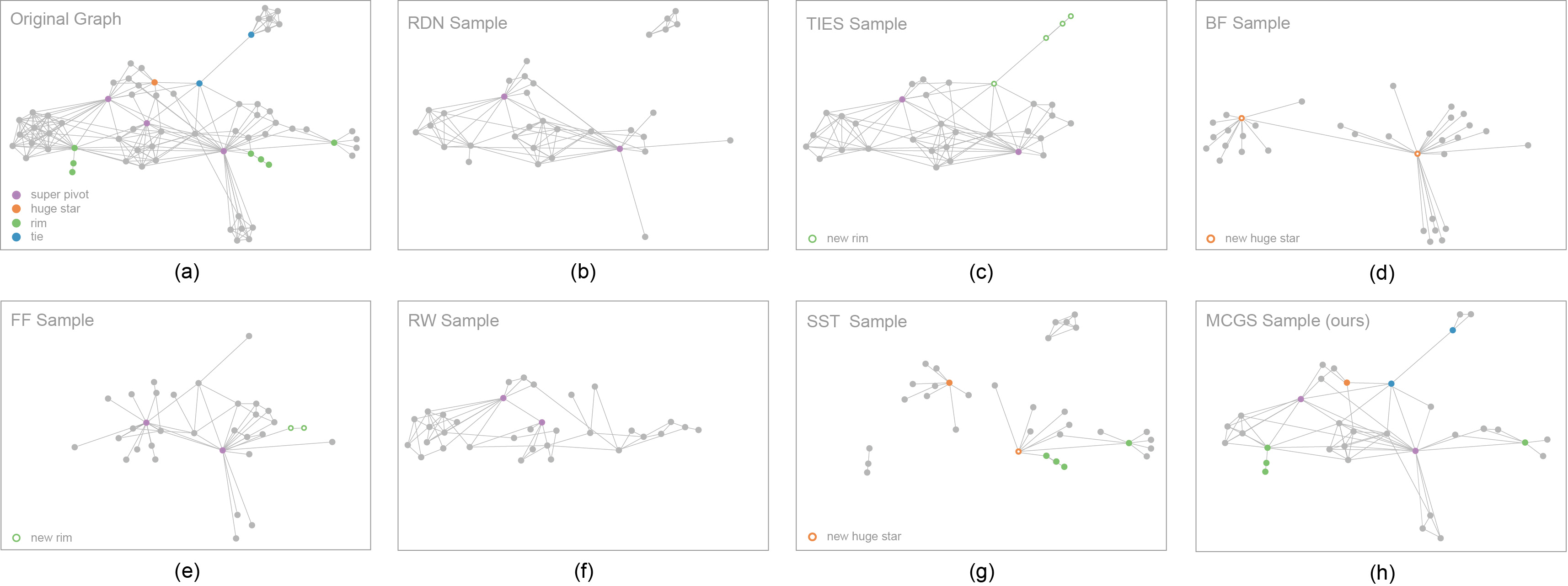}
	\caption{Illustration of four representative types of minority structures in a toy case graph (a) and seven graph samples obtained by RDN (b), TIES (c), BF (d), FF (e), RW (f), SST (g), and our proposed MCGS (h), respectively, with a sampling rate of 50$\%$. The graph is slightly modified from the character relationship network of the novel $\emph{Les Misérables}$ \cite{A14}. The relative locations of nodes in a sample are consistent with those of the corresponding nodes in the original graph. Solid-color dots represent original minority structures. Hollow-color dots represent newly generated minority structures.}
	\label{fig:figure1}
\end{figure*}

{To verify this hypothesis, we conducted a pilot user study and an experimental study. In the user study (Section \ref{sec:sec3}), we recruited 20 participants and asked them to freely select structures of interest (SOIs) in 34 real-world graph data sets. The results showed that four representative types of minority structures, namely, super pivot, huge star, rim, and tie, elicited strong interest from the participants. Super pivots and huge stars are a small proportion of nodes with extremely high degrees. Rims are parachute- or chain-like structures attached to community margins. Ties are sparely distributed bridges at community boundaries. \autoref{fig:figure1}(a) shows a toy case graph containing three super pivots, one huge star, three rims, and one tie.}

%\vspace{-10pt}
\par{In the experimental study (Section \ref{sec:sec4}), we selected 12 real-world graph data sets and 20 reference graph sampling algorithms and designed three new quantitative indicators to evaluate their performance on preserving the four types of minority structures. The results showed that most of the algorithms cannot effectively preserve minority structures and may generate new minority structures that did not exist in the original graphs, especially for huge stars, rims, and ties. \autoref{fig:figure1}(b$  -  $g) show the samples obtained by popular graph sampling algorithms. Most of the samples fail to preserve the huge stars, rims, and ties in \autoref{fig:figure1}(a). New huge stars or new rims are found in \autoref{fig:figure1}(c, d, e, and g). This situation is detrimental to graph analyses that focus on minority structures.}
%\vspace{-5pt}

\par{A new graph sampling algorithm oriented to minority structure analysis is needed, but its design is difficult. On the basis of results and experience in the experimental study, five key points should be seriously considered in the design: (1) identifying minority structures quickly and accurately; (2) avoiding losing minority structures in samples caused by the undersampling of their neighbors \cite{A12}, such as the loss of the parachute-like rim in \autoref{fig:figure1}(b, d, and e); (3) minimizing the inconsistency of preserved minority structures in random sampling; (4) balancing the preservation between minority and majority structures; and (5) suppressing the generation of new minority structures.}
%\vspace{-5pt}

\par{We propose a new graph sampling method called mino-centric graph sampling (MCGS) by considering the above points. {This work stipulates graphs are simple, unattributed, undirected, and connected to simplify representations.} First, we design a fast triangle-based algorithm to identify super pivots and huge stars and a fast cut-point-based algorithm to identify rims and ties. Second, we propose a set of importance assessment criteria to guide the preservation of minority structures and their neighbors and minimize the influence of random sampling. Finally, we introduce three optimization objectives into a greedy strategy to strike a balance between the preservation of minority and majority structures and suppress the generation of new minority structures. A series of experiments and case studies are conducted to evaluate the effectiveness of the proposed MCGS. The results reveal that MCGS performs the best among the 20 reference algorithms on the preservation of minority structures and suppression of new minority structure generation. MCGS also achieves highly satisfactory performance on the majority structure preservation.}
%\vspace{-5pt}

\par{In summary, this work presents the first attempt to investigate minority structure preservation in graph sampling. This work contributes: (1) four representative types of minority structures that are summarized through a controlled user study, (2) an experimental study that evaluates the performance of existing graph sampling algorithms on minority structure preservation, and (3) a new graph sampling algorithm that is oriented to minority structure preservation.}

%%2 %%%%%%%%%%%%%
\section{Related Work}
%2.1
\subsection{Graph Structure Analysis}
\label{sec:sec2.1}
Graph structures have been categorized and defined from diverse analytical perspectives. In this work, we categorize graph structures into minority and majority and define four types of minority structures (i.e., super pivot, huge star, rim, and tie). Our categorization and definition have certain connections with those in graph anomaly detection, community detection \cite{G31}, structural role discovery \cite{B52}, and frequent subgraph mining \cite{B43}.
%modern graph science \cite{B1,B2}. namely, communities \cite{B3}, motifs \cite{B20}, and anomalies \cite{A2}

\par{Graph structures are categorized into normal and abnormal in graph anomaly detection. Anomalies are nodes, edges, or subgraphs that differ from most ones \cite{B10,B11}. In general, node and edge anomalies are minority structures, such as spammers in email networks with extremely high degrees, because of low occurring probability and small size. However, subgraph anomalies with relatively large sizes are not minority structures.}
%\par{Anomalies are graph objects that differ from most objects in a graph \cite{B10,B11}, such as spammers in email networks with extremely high degrees \cite{B12} and neutral voters in voting networks with uncertain voting orientations \cite{B13}. Many techniques have been developed for graph anomaly detection, such as Oddball \cite{B14}, Fbox \cite{B15}, and SUBDUE \cite{B16}. In general, anomalies, such as extremely high degree nodes \cite{B14,B44}, are considered minority structures because of their low occurring probability.}

\par{Graph structures can be categorized into communities/clusters, hubs, and outliers to facilitate community detection \cite{B65}. Communities are majority structures \cite{B50}. Hubs correspond to ties in minority structures. Outliers refer to small structures attaching at margins of communities, such as secret leaders controlling a criminal gang through intermediaries \cite{B60}, corresponding to chain-like rims in minority structures.}
%\par{Communities are groups of densely interconnected nodes \cite{B5,B6}, and many featured methods have been developed for community detection, such as fast unfolding \cite{B7}, spectral clustering \cite{B8}, and non-negative matrix factorization \cite{B9}. In this study, communities are regarded as majority structures because they generally involve a certain number of nodes. As minority structures, ties and rims coexist with communities and appear in community boundaries and margins.}
%\par{Motifs are basic structural units of graphs \cite{B17,B19}, such as triangles, quadrangles, and stars, which frequently appear and represent a series of duplicate features. Studying motifs is of particular interest in many domains \cite{B4}. In system biology, some motifs act as functional units that ensure cell regulatory processes \cite{B18}. Lacroix \cite{B21} proposed a method to count such motifs. In visualization and visual analytics, motifs can be used to optimize node-link diagrams. Wang et al. \cite{B30} applied motif-shape constraints in graph layouts, Bach et al. \cite{B19} designed a motif-oriented edge bundling method, and Kwon et al. \cite{B22} used motif-frequencies to measure graph layout similarity. In the present study, motifs are not considered as typical minority structures because of their high occurring probability. However, some variations of motifs are minority structures, such as huge stars.}

\par{Structural role discovery assigns behavior roles, such as clique- or periphery-members, to structures \cite{B51,B52}. Cliques are not minority structures because of containing many densely connected nodes, but extremely high degree nodes in cliques are minority structures. Some periphery members at clique marginal areas, such as parachute- and chain-like nodes, are rims in minority structures.}

\par{In the fields of frequent subgraph mining \cite{B43}, motifs discovery \cite{B53}, and graphlet-based characterization \cite{B62,B63}, graph structures are categorized depending on occurrence frequency. Most frequent subgraphs, motifs, and graphlets are majority structures. However, their variants could be minority structures. For example, large star-shaped motifs can be regarded as huge stars in minority structures but only the central high degree nodes of such motifs are included in our definition of huge stars. Moreover, many studies of attributed data analysis devoted to rare category detection \cite{B55} and classified data entities into majority and minority classes \cite{B57}, which supports this work.}

%\par{The rare category refers to rare but feature-similar data entities that are obscured within a large but feature-different data set \cite{B58}. Such data entities are very important in many applications, such as financial fraud detection and rare disease discovery \cite{B64}.}
%B58 B64 删除

%2.2
\subsection{Graph Sampling Algorithms}
\label{sec:sec2.2}
Existing graph sampling algorithms can be classified into three groups, namely, node-, edge-, and traversal-based \cite{A4,A12}. In the node-based and edge-based groups, Random Node (RN), Random PageRank Node (RPN), and Random Degree Node (RDN) \cite{A8} are typical node-based algorithms. Random Edge (RE) \cite{A8} and Random Node Edge (RNE) \cite{B23} are classic edge-based algorithms. These algorithms are lightweight and provide theoretical references and building blocks for advanced algorithms. However, these algorithms have a common defect that randomly selected nodes are uncorrelated, thus causing unsatisfying preservation of graph connectivity \cite{A8}. Totally-Induced Edge Sampling (TIES) \cite{B24} is an edge-based approach that introduces a graph induction step into sampling. Additional edges in this step are retained to restore graph connectivity. We learn from this idea to reduce the generation of new minority structures.

\par{The traversal-based group includes many algorithms. Their common merit is that connected graphs remain connected after sampling \cite{B26}. Breadth-First (BF) and Depth-First (DF) samplings are two basic approaches that select nodes in a breadth-first and depth-first traversal order, respectively \cite{B26,B25}. Snowball (SB) and Forest Fire (FF) are variants of BF \cite{B62} that expand limited neighbors instead of exhaustive expansion to ensure the picking of depth nodes \cite{A4,A8}. Random Walk (RW) and Random Jump (RJ) are variants of DF \cite{B29,B27} that allow walking to neighbors or random nodes during the traversal for preserving the global topology. Variants of RW include Multi-Dimensional Random Walk (MDRW) \cite{B32}, Metropolis–Hastings Random Walk (MHRW) \cite{B34}, Rejection-Controlled Metropolis-Hastings (RCMH) \cite{B37}, and Generalized Maximum-Degree Random Walk (GMD) \cite{B35}. In our approach, a depth-first traversal and random-pick processing are adopted.}

\par{Some graph sampling approaches cannot be directly sorted into the above three groups. These approaches provide alternative solutions for specific goals. Random Areas Selection Sampling (RAS) selects an area of nodes each time to fully preserve their neighborhood structures \cite{B43}. We use this idea to preserve neighbors of minority structures. Distributed Learning Automata Sampling (DLAS) uses multiple automata for sampling \cite{B36}. Sampling with Shortest Paths (SSP) and Sampling with Spanning Trees (SST) identify important edges to guide sampling \cite{B38,B39}. Multiple Snowball with Cohen (RMSC) combines the advantages of RN and SB. Sampling based on Graph Partition (SGP) and Sampling based on Densification Power Law (DPL) partition a graph before sampling \cite{B40,B41}. We adopt this partition idea to deal with irregularly shaped graphs. {Moreover, all the aforementioned algorithms perform graph sampling in the data space rather than the visual space to simultaneously support graph computation acceleration and graph visualization simplification. Our MCGS is also a data space approach.}}

%2.3
\subsection{Evaluation of Graph Sampling}
\label{sec:sec2.3}
Graph sampling algorithms have been thoroughly evaluated by two groups of metrics. The first group quantifies how well structural properties of the original graph are preserved \cite{A7,A8,B32}. Popular metrics are the Degree Distribution (DD) and Clustering Coefficient Distribution (CCD) \cite{B46,B47}. The second group measures the similarity between the original and sampled graphs. Two popular metrics in this group are the Jaccard Index (JI) that measures the similarity by the size of intersections \cite{E9} and the number of connected components (NCC) that measures the similarity of graph connectivity \cite{G32}. Recently, visual perception factors are considered in graph sampling evaluation. Wu et al. \cite{A12} found that three factors, namely, cluster quality, high degree nodes, and coverage area, influence the visual perception of sampled node-link diagrams. Quan et al. \cite{B42} studied proxy graphs to measure the shape-based faithfulness of sampled graphs. 

\par{At present, no metrics are tailored for evaluating the preservation of majority and minority structures. In this work, we use traditional metrics, including DD, JI, and NCC, to evaluate majority structure preservation because inherent connections exist between these metrics and structural properties or overall shapes of majority structures. For example, DD can measure the structural properties of frequent structures. JI can compare the overall shapes formed by large structures. Moreover, we design three new indicators to evaluate minority structure preservation. We also consider perception factors in our evaluation experiments \cite{addEvaluating,addMulti,addLdsscanner,addSurvey}.}

%%3 %%%%%%%%%%%%%
\section{Pilot User Study}
\label{sec:sec3}
We assumed that existing graph sampling algorithms cannot effectively preserve minority structures. Before verifying this hypothesis, we conducted a pilot user study to answer two basic guiding questions: whether and which minority structures are important in graph analysis.

%3.1
\subsection{User Study Design}
We recruited 20 participants (8 females and 12 males, all were graduate students aged 20-26 years) and selected 34 real-world graph data sets. The task was to select any SOIs in the graphs. The graphs were visualized in node-link diagrams as plain graphs. Graph layouts used the ForceAtlas2 algorithm \cite{G39}. The participants were asked to perform the task on the 34 graphs in a random order. For a graph, 60 s was allotted, and an interval of 2 s was set before proceeding to the next graph. After completing all graphs, the participants reviewed their selections and stated their thoughts. The study was conducted on a desktop with a 23.8-inch 1920 × 1080 LCD display, a standard keyboard, and a mouse. {Descriptions of data sets and the experimental interface are provided in the supplementary material.}

%3.2
\subsection{Result Analysis}
The selected SOIs and selection sequence of each participant on each graph were recorded as the results. We manually categorized all SOIs into eight types and counted the entries for each type. If multiple SOIs of the same type were sequentially selected by a participant in a graph, then the count of the type was only 1 to avoid overcounting. We also counted the entries for each type in orders of 1st, 2nd, 3rd, and others in all sequences. \autoref{tab:table1} shows the statistical results of the eight SOI types.

\begin{table}[!htbp]
	%\tiny
	\scriptsize
	%fontsize{5pt}{1.5}
	\renewcommand\tabcolsep{10.7pt}  %% 调整表格列间的宽度
	\renewcommand\arraystretch{1.2} %设置表格行间距
	\centering
	\vspace{-0.15cm}
	%\vspace{-0.05cm}  %调整图片与上文的垂直距离
	%\setlength{\abovecaptionskip}{-0.03cm}   %调整图片标题与图距离	
	%\setlength{\belowcaptionskip}{-0.5cm}   %调整图片标题与下文距离
	\caption{Statistics of eight SOI types for the participants in the user study}
	\label{tab:table1}
	%\linespread{1. 0}\selectfont

	\begin{tabular}{|p{1.65cm}<{\centering}|p{0.56cm}<{\centering}|p{0.56cm}<{\centering}|p{0.57cm}<{\centering}|p{0.66cm}<{\centering}|p{0.58cm}<{\centering}|}
	%\begin{tabular}{|c|c|c|c|c|c|}%一个c表示有一列，格式为居中显示(center)
		\hline  % 在表格最上方绘制横线
		\multirow{2}{*}{\textbf{SOI Type}}&
		\multicolumn{5}{c|}{\textbf{Number of Entries by Order}}\\
		\cline{2-6}
					
		&{\textbf{1st}}&{\textbf{2nd}}&{\textbf{3rd}}&{\textls[-30]{\textbf{Others}}}&{\textls[-25]{\textbf{Total}}}\\
		%&{\textbf{1st}}&{\textbf{2nd}}&{\textbf{3rd}}&{\textbf{Others}}&{\textbf{Total}}\\
		\cline{1-6}		
		{\textls[-20]{HD-global}}&184&146&72&24&426\\
		\cline{1-6}		
		{{MS}}&103&124&103&51&381\\
		\cline{1-6}		
		{{BS}}&87&81&128&49&345\\
		\cline{1-6}		
		{{CS}}&22&65&58&84&229\\
		\cline{1-6}	
		{{FC}}&41&18&14&22&95\\
		\cline{1-6}	
		{\textls[-35]{CS-overlapping}}&15&15&14&17&61\\
		%\hline
		\cline{1-6}	
		{\textls[-20]{HD-local}}&4&4&20&4&32\\
		\hline
		{{IS}}&10&1&3&4&18\\
		\cline{1-6}
		%\hline
	\end{tabular}
	\vspace{-0.22cm}
\end{table}

\par{The result showed that four SOI types, namely, global high degree structure (HD-global), margin structure (MS), boundary structure (BS), and community structure (CS), had entries more than the average of 198 for all types. The remaining four SOI types, namely, small cliques far away from graph main bodies (FC), community overlapping structure (CS-overlapping), local high degree structure (HD-local), and isolated structure (IS), had entries far fewer than the average. Therefore, HD-global, MS, BS, and CS were considered as the most appealing structures in interactive graph explorations. Among them, HD-global, MS, and BS belonged to minority structures, whereas CS belonged to majority structures.}

\par{HD-global, MS, and BS must be further discussed. (1) HD-global ranked first. Most of the participants confirmed that the nodes with extremely high degrees had a strong visual saliency, which was in line with the previous research that stated that visually salient high degree nodes should not be lower than the global top 10$\%$ \cite{C3}. In general, high degree nodes have two subtypes \cite{D3}, namely, pivot and star. A pivot is a high degree node whose neighbors have at least one interconnection. A star is a high degree node whose neighbors do not interconnect. We stipulated that our concerned pivots had degrees within the global top 5$\%$, named as super pivots; our concerned stars had degrees above the global mean \cite{D1}, named as huge stars. We used a strict threshold for pivots but a lax one for stars because stars were lower in quantity than pivots. (2) MS structures ranked second. They were appendage nodes that occasionally occurred in the marginal areas of communities and formed specific visual shapes, such as shapes like parachute, chain, balloon, or tree. A large proportion of the participants reported that parachute- and chain-like structures at community rims were especially eye-catching. Thus, we used rims as a concrete representative of MS structures with the two visual shapes. (3) BS structures ranked third and were a sequence of nodes bridging any two communities. Many of the participants commented that the structures that tied communities were sometimes more attractive than communities. Thus, we used ties as a concrete representative of BS structures in this work.}

\par{{As a result, we obtained four representative types of minority structures (i.e., super pivot, huge star, rim, and tie). These structures were visually salient in node-link diagrams and elicited strong interest from the participants in the pilot user study. A literature review (Section \ref{sec:sec2.1}) confirmed that these structures were also important in various research and application branches.}}

%%4 %%%%%%%%%%%%%
\section{Experimental Study}
\label{sec:sec4}
We conducted a controlled experiment to examine the performance of existing graph sampling algorithms in preserving the four representative types of minority structures.
%4.1
\subsection{Hypotheses}
\label{sec:sec4.1}
To guide the experiment, we formulated three specific hypotheses:

\par{\textbf{H1:} Existing algorithms pursue graph similarity and have a low ability to preserve minority structures. At present, graph similarity is largely measured by the structural properties and overall shapes of majority structures (Section \ref{sec:sec2.3}). Thus, we assume that existing algorithms naturally have a relatively low ability for minority structure preservation.}

\par{\textbf{H2:} Existing algorithms may produce new minority structures. Many nodes in samples inevitably have incomplete original neighbors, thereby causing that some structures may degenerate to minority structures that do not exist in the original graph.}

\par{\textbf{H3:} Existing algorithms cannot guarantee the preservation of important minority structures. Only a part of minority structures in a graph are crucially important according to certain criteria (Section \ref{sec:sec5.4.2}). The randomness of graph sampling has a fatal influence on minority structure preservation. Slight differences in selecting nodes may lead to the disappearance of minority structures. Thus, we assume that important minority structures will disappear or be no longer important in samples.}

%4.2
\subsection{Experimental Study Design}
\textbf{Data preparation.} We selected 12 real-world graph data sets as the experiment data, two of which were for pilot tests to determine the design of the formal experiment. The graphs were mainly social, web, and communication networks popular in graph studies and included seven small, three medium, and two large scales. Data processing was conducted to detect and label the type and importance of each minority structure in each graph by using the methods introduced in Sections \ref{sec:sec5.4.1} and \ref{sec:sec5.4.2}. 

\par{\textbf{Reference algorithms and parameter settings.} We selected 20 graph sampling algorithms as references. These algorithms had two common parameters, namely, initial seed and sampling rate. To reduce the influence of parameter settings, we prepared four types of seeds \cite{D1} (i.e., random, high degree, high betweenness, and peripheral nodes) and four sampling rates (i.e., 10$\%$, 20$\%$, 30$\%$, and 40$\%$). Other distinctive parameters were set with defaults. Detailed data descriptions and parameter setting considerations are provided in the supplementary material.}

\par{\textbf{Experimental Procedure.} Each algorithm ran 800 trials (10 graphs × 4 types of seeds × 4 different sampling rates × 5 runs). 16,000 samples in total were obtained as the raw results. The experiment was conducted on a desktop with a 3.4 GHz Intel i7 CPU and 16 GB of RAM.}

%4.3
\subsection{Indicator Design}
\label{sec:sec4.3}
{Given the lack of established indicators, we designed three quantitative indicators to verify the three hypotheses (Section \ref{sec:sec4.1}) and measure the performance of minority structure preservation.}

\par{\textbf{Minority structure preservation rate (MSPR).} This indicator is the ratio of the preservation rate of minority structures to a sampling rate (H1), in which the preservation rate refers to the proportion of original minority structures preserved in a sample. For example, given a sample with a sampling rate of 30$\%$, the MSPR is 1 when 3 out of 10 minority structures in the original graph are preserved. Generally, MSPR is related to a certain type of minority structure and is defined as

\vspace{-0.1cm}
\begin{equation}
\setlength{\abovedisplayskip}{4.5pt}
\setlength{\belowdisplayskip}{4.5pt}
MSPR=\frac{\mid MS_{S_{x}}\cap MS_{x}\mid/\mid MS_{x}\mid}{\Phi},
\nonumber
\end{equation}

\noindent{where $MS_{x}$ represents the set of minority structures of \textit{x} type in a graph \textit{G}; $MS_{S_{x}}$ represents the set of minority structures of \textit{x} type in a sample $G_{s}$; $\Phi$ is the sampling rate; and $\mid MS_{x}\mid$ and $\mid MS_{S_{x}}\mid$ are the cardinalities of $MS_{x}$ and $MS_{S_{x}}$ respectively. An MSPR approaching, equal to, or even greater than 1 means a “perfect” result.}}

\par{\textbf{New minority structure generation rate (MSGR).} This indicator represents the probability that new minority structures of a certain type occur in a sample (H2). An MSGR approaching or equaling 0 is a “perfect” result. MSGR is formulated as}

\vspace{-0.1cm}
\begin{equation}
\setlength{\abovedisplayskip}{4.5pt}
\setlength{\belowdisplayskip}{4.5pt}
MSGR=\mid \left (MS_{S_{x}}-MS_{x}\right )\mid/\mid MS_{S_{x}}\mid.
\nonumber
\end{equation}

\par{\textbf{Mean importance precision (MIP).} This indicator evaluates the mean preservation precision of top $K$ minority structures of a certain type before and after sampling (H3). MIP is a variant of average precision in recommendation system ranking \cite{D2}. We define MIP as

\vspace{-0.1cm}
\begin{equation}
\setlength{\abovedisplayskip}{4.5pt}
\setlength{\belowdisplayskip}{4.5pt}
MIP=\frac{\displaystyle\sum_{i=1}^{K} \mid Top_{i}(MS_{x})\cap Top_{i}(MS_{S_{x}})\mid /i}{K},
\nonumber
\end{equation}

\noindent{where $Top_i(.)$ denotes the top \textit{i} important minority structures. For example, the top five super pivots in a graph are nodes ID-11, 12, 26, 9, and 15; and those in a sample are nodes ID-11, 18, 26, 12, and 15. Four super pivots at the 1st, 3rd, 4th, and 5th positions in the sample are matched. Therefore, MIP is 0.743 [(1/1+1/2+2/3+3/4+4/5)/5]. An MIP approaching or equal to 1 is a “perfect” result.}}

%4.4
\subsection{Result Analysis}
\textbf{Result processing.} The result processing consisted of three parts. (1) We calculated the values of the three indicators for the four types of minority structures on each sample. (2) We calculated the medians and standard deviations of each indicator on each minority structure type and algorithm. {(3) We selected empirical thresholds based on two criteria: indicating good sampling results and differentiating the performance of the reference algorithms. We stipulated that MSPR $\geq$ 0.9 was good results, indicating that the preservation rate of minority structures was approximately equal to the sampling rate. Empirically, this threshold represents an ideal balance for the preservation of minority and majority structures. MSGR $\leq$ 0.5 denoted good results, representing that new minority structures were no more in quantity than the preserved original ones. Thus, original minority structures can still be dominant in the sample and analyzed without distinct interference. MIP $\geq$ 0.5 was good, implying that more than half of the top \textit{K} important minority structures were preserved.}

\par{We verified the hypotheses based on the processed results (\autoref{tab:table2}). Additional results are provided in the supplementary material.}

\par{\textbf{Hypothesis verification.} H1 was partially confirmed. The MSPR results reflected that these algorithms generally had low ability in preserving minority structures but a few algorithms can effectively preserve a certain minority structure type. For the four types of minority structures, the MSPR results of super pivots ($\mu$ = 0.6733) and huge stars ($\mu$ = 0.4387) were better than those of rims ($\mu$ = 0.2835) and ties ($\mu$ = 0.1199). The reason was that the former two were commonly embedded in communities with relatively stable neighborhood structures, whereas the latter two were located at the margin or boundary areas of communities with sparse and unstable neighborhoods. From a single-algorithm perspective, RDN, RPN, DPL, and RAS performed well (MSPR $\geq$ 0.9) in super pivots because they were in favor of high degree nodes \cite{A8,B38,B37}. TIES also performed well in super pivots because of its graph induction step \cite{B24}. SST performed well in rims because the use of spanning trees maintained peripheral nodes \cite{B39}.}

\par{H2 was partially confirmed. The MSGR results reflected that these algorithms can effectively suppress the generation of new super pivots ($\mu$ = 0.3212) but hardly inhibited the generation of new huge stars ($\mu$ = 0.6304), rims ($\mu$ = 0.8231), and ties ($\mu$ = 0.6947). Many algorithms performed well (MSGR $\leq$ 0.5) in super pivots because other structures rarely degenerated to pivots after sampling. However, pivots may degenerate to stars when all edges between neighbors were lost. Thus, only seven algorithms performed well in huge stars. Furthermore, peripheral nodes and CS-overlapping structures had chances to degenerate to rims and ties, respectively. RMSC was the only algorithm that effectively suppressed the generation of new ties. Its breadth-first and multi-snowball strategy effectively preserved the connections between communities \cite{B37}.}

\par{H3 was fully confirmed. Most of the MIP medians in \autoref{tab:table2} were poor (lower than 0.5), indicating that no algorithms can guarantee the preservation of important minority structures for two reasons. First, originally important minority structures were not well preserved. Second, newly generated minority structures became important in samples.}

\par{\textbf{Other findings.} The graph data sets, sampling rates, and initial seeds affected minority structure preservation to some extent. (1) Data sets. A graph with a single cluster or multiple clusters in approximate sizes is called a balanced graph. A graph with multiple clusters that present a wide difference in size is called an unbalanced graph. We found that the results on unbalanced graphs were worse than those on balanced graphs, because small clusters that contained important ties or chain-like rims in unbalanced graphs were not effectively maintained. (2) Sampling rates. The scores of the three indicators generally improved when the sampling rate increased because high sampling rates resulted in highly completed structures. (3) Initial seeds. The high-degree type of initial seeds generally performed better than the other three types of initial seeds because the former provided numerous available paths for sampling. }
%{table}有若干可选参数 [!htbp]
%h代表here,将表格排在当前文字位置
%t 表示将表格放在下一页的 top (页首)
%b 表示将表格放在当前页的 bottom (底部)
%!表示忽略美观因素，尽可能按照参数指定的方式来处理表格浮动位置。
%表格将会按照所给参数，依次尝试按照每个参数进行排版，当无法排版时，将会按照下一个参数
\begin{table*}[!htbp]
	\tiny
	\scriptsize
	%\fontsize{6.5pt}{1.5}
	\renewcommand\tabcolsep{4.5pt}  %% 调整表格列间的宽度
	\renewcommand\arraystretch{1.15} %设置表格行间距
	\centering
	%\vspace{-0.05cm}  %调整图片与上文的垂直距离
	%\setlength{\abovecaptionskip}{-0.03cm}   %调整图片标题与图距离	
	\setlength{\belowcaptionskip}{-0.2cm}   %调整图片标题与下文距离
	\caption{Results of the experiments in Sections \ref{sec:sec4}, \ref{sec:sec6.1.1}, and \ref{sec:sec6.1.2} in terms of the medians of indicators (columns) and algorithms (rows). \textit{P}, \textit{S}, \textit{R}, and \textit{T} represent super pivot, huge star, rim, and tie, respectively. Blue indicates the winners in significance tests among algorithms. Bold indicates that the value is better than the empirical “good” indicator threshold (only for MSPR, MSGR, and MIP). Using the first column as an example, significant differences are found among algorithms on MSPR and super pivot. RDN, RPN, TIES, and MCGS are the winners. Six algorithms obtain “good” results of MSPR on super  pivot.}
	\label{tab:table2}
	\newcommand{\tabincell}[2]{\begin{tabular}{@{}#1@{}}#2\end{tabular}}
%	\begin{tabular}{|m{1.4cm}<{\centering}|m{1.4cm}<{\centering}|m{0.6cm}<{\centering}|
%			m{0.6cm}<{\centering}|m{0.6cm}<{\centering}|m{0.6cm}<{\centering}|
%			m{0.6cm}<{\centering}|m{0.6cm}<{\centering}|m{0.6cm}<{\centering}|
%			m{0.6cm}<{\centering}|m{0.6cm}<{\centering}|m{0.6cm}<{\centering}|
%			m{0.6cm}<{\centering}|m{0.6cm}<{\centering}|m{0.6cm}<{\centering}|
%			m{0.6cm}<{\centering}|m{0.6cm}<{\centering}|m{0.6cm}<{\centering}|}
	\begin{tabular}{|m{1.3cm}<{\centering}|m{1.58cm}<{\centering}|c|c|c|c|c|c|c|c|c|c|c|c|c|c|c|c|}%一个c表示有一列，格式为居中显示(center)
		\hline  % 在表格最上方绘制横线
		\rule{0pt}{4pt}\multirow{3}*{\textbf{\tabincell{c}{Sampling \\ Strategy}}}&
		\multirow{3}*{\textbf{Algorithm}}&
		\multicolumn{12}{c|}{\textbf{Indicators for Minority Structure Preservation}}&
		%\vspace{-2pt}
		\multicolumn{4}{c|}{\textbf{\tabincell{c}{\vspace{-3pt}{Indicators for Majority} \\ {Structure Preservation}}}}\\
		\cline{3-18}
			
		&&\multicolumn{4}{c|}{\textbf{MSPR}}&
		\multicolumn{4}{c|}{\textbf{MSGR}}&
		\multicolumn{4}{c|}{\textbf{MIP}}&
		\multirow{2}{*}{\textbf{KSD}}&
		\multirow{2}{*}{\textbf{SDD}}&
		\multirow{2}{*}{\textbf{RCC}}&
		\multirow{2}{*}{\textbf{JI}}\\
		\cline{3-14}
		
		&&\textbf{\textit{P}} &\textbf{\textit{S}} &\textbf{\textit{R}} &\textbf{\textit{T}}
		&\textbf{\textit{P}} &\textbf{\textit{S}} &\textbf{\textit{R}} &\textbf{\textit{T}}
		&\textbf{\textit{P}} &\textbf{\textit{S}} &\textbf{\textit{R}} &\textbf{\textit{T}}&&&& \\		\hline
		
		\multirow{3}{*}{{Node-based}}&
		{RN}&
		0.796&0.453&0.098&0.000&
		\textbf{0.238}&0.723&0.924&1.000&
		0.000&0.000&0.000&0.000&
		0.495&0.080&0.006&0.066\\
		\hhline{|~|-|-|-|-|-|-|-|-|-|-|-|-|-|-|-|-|-|}
		
		&{RDN}&
		\cellcolor{LightBlue2}\textbf{0.997}&0.015&0.000&0.000&
		\cellcolor{LightBlue2}\textbf{0.000}&\cellcolor{LightBlue2}\textbf{0.000}&1.000&\cellcolor{LightBlue2}\textbf{0.479}&
		0.457&0.000&0.000&0.000&
		0.224&0.003&0.167&0.152\\
		\hhline{|~|-|-|-|-|-|-|-|-|-|-|-|-|-|-|-|-|-|}
		
		&{RPN}&
		\cellcolor{LightBlue2}\textbf{0.994}&0.256&0.017&0.000&
		\cellcolor{LightBlue2}\textbf{0.000}&\cellcolor{LightBlue2}\textbf{0.000}&0.948&0.854&
		0.457&0.000&0.000&0.000&
		0.222&\cellcolor{LightBlue2}0.003&0.067&0.128\\
		\hline
		
		\multirow{2}{*}{{Edge-based}}&
		{RNE}&
		0.704&0.346&0.135&0.000&
		\textbf{0.295}&1.000&0.851&1.000&
		0.000&0.000&0.000&0.000&
		0.353&0.006&\cellcolor{LightBlue2}1.000&0.057\\
		\hhline{|~|-|-|-|-|-|-|-|-|-|-|-|-|-|-|-|-|-|}
		
		&{TIES}&
		\cellcolor{LightBlue2}\textbf{0.994}&0.172&0.083&0.000&
		\cellcolor{LightBlue2}\textbf{0.000}&\cellcolor{LightBlue2}\textbf{0.000}&0.873&0.889&
		0.457&0.000&0.000&0.000&
		\cellcolor{LightBlue2}0.182&\cellcolor{LightBlue2}0.003&0.333&0.144\\
		\hline
		
		\multirow{8}{*}{{\tabincell{c}{Traversal-\\based}}}&
		{BF}&
		0.418&0.000&0.007&0.000&
		0.580&1.000&0.998&1.000&
		0.000&0.000&0.000&0.000&
		0.713&0.012&\cellcolor{LightBlue2}1.000&0.036\\
		\hhline{|~|-|-|-|-|-|-|-|-|-|-|-|-|-|-|-|-|-|}
		
		&{FF}&
		0.691&0.199&0.044&0.000&
		\textbf{0.308}&0.938&0.985&\cellcolor{LightBlue2}0.729&
		0.000&0.000&0.000&0.000&
		0.424&0.003&\cellcolor{LightBlue2}1.000&0.050\\
		\hhline{|~|-|-|-|-|-|-|-|-|-|-|-|-|-|-|-|-|-|}
		
		&{GMD}&
		0.705&0.350&0.131&0.000&
		\textbf{0.293}&1.000&0.851&1.000&
		0.000&0.000&0.000&0.000&
		0.354&0.006&\cellcolor{LightBlue2}1.000&0.056\\
		\hhline{|~|-|-|-|-|-|-|-|-|-|-|-|-|-|-|-|-|-|}
		
		&{MDRW}&
		0.700&0.350&0.113&0.000&
		\textbf{0.303}&1.000&0.875&1.000&
		0.000&0.000&0.000&0.000&
		0.373&0.006&\cellcolor{LightBlue2}1.000&0.051\\
		\hhline{|~|-|-|-|-|-|-|-|-|-|-|-|-|-|-|-|-|-|}
		
		&{MHRW}&
		0.264&0.035&0.092&0.000&
		0.750&0.948&0.945&1.000&
		0.000&0.000&0.000&0.000&
		0.662&0.014&0.015&0.046\\
		\hhline{|~|-|-|-|-|-|-|-|-|-|-|-|-|-|-|-|-|-|}
		
		&{RCMH}&
		0.496&0.603&0.174&0.000&
		0.526&0.991&0.818&0.963&
		0.000&0.000&0.000&0.000&
		0.391&0.006&\cellcolor{LightBlue2}1.000&0.059\\
		\hhline{|~|-|-|-|-|-|-|-|-|-|-|-|-|-|-|-|-|-|}
		
		&{RJ}&
		0.677&0.377&0.119&0.000&
		\textbf{0.316}&1.000&0.903&1.000&
		0.000&0.000&0.000&0.000&
		0.351&\cellcolor{LightBlue2}0.006&\cellcolor{LightBlue2}1.000&0.057\\
		\hhline{|~|-|-|-|-|-|-|-|-|-|-|-|-|-|-|-|-|-|}
		
		&{RW}&
		0.713&0.355&0.125&0.000&
		\textbf{0.286}&1.000&0.857&1.000&
		0.000&0.000&0.000&0.000&
		0.363&0.006&\cellcolor{LightBlue2}1.000&0.056\\
		\hline
		
		\multirow{8}{*}{{Others}}&
		{DLAS}&
		0.734&0.421&0.093&0.000&
		\textbf{0.265}&\textbf{0.337}&0.861&0.952&
		0.000&0.000&0.000&0.000&
		0.359&0.004&0.017&0.114\\
		\hhline{|~|-|-|-|-|-|-|-|-|-|-|-|-|-|-|-|-|-|}
		
		&{DPL}&
		\textbf{0.946}&0.297&0.000&0.000&
		\textbf{0.019}&\textbf{0.250}&1.000&1.000&
		0.000&0.000&0.000&0.000&
		0.254&\cellcolor{LightBlue2}0.004&0.083&0.098\\
		\hhline{|~|-|-|-|-|-|-|-|-|-|-|-|-|-|-|-|-|-|}
		
		&{RAS}&
		\textbf{0.962}&0.113&0.131&0.000&
		\textbf{0.000}&\cellcolor{LightBlue2}\textbf{0.000}&0.611&\cellcolor{LightBlue2}0.633&
		0.457&0.000&0.000&0.000&
		\cellcolor{LightBlue2}0.181&\cellcolor{LightBlue2}0.003&0.500&0.157\\
		\hhline{|~|-|-|-|-|-|-|-|-|-|-|-|-|-|-|-|-|-|}
		
		&{RMSC}&
		0.432&0.000&0.000&0.000&
		0.548&0.989&0.945&\cellcolor{LightBlue2}\textbf{0.000}&
		0.000&0.000&0.000&0.000&
		0.989&0.017&0.003&0.000\\
		\hhline{|~|-|-|-|-|-|-|-|-|-|-|-|-|-|-|-|-|-|}
		
		&{SGP}&
		0.877&0.000&0.034&0.000&
		\textbf{0.125}&\textbf{0.425}&0.891&1.000&
		0.000&0.000&0.000&0.000&
		0.305&\cellcolor{LightBlue2}0.004&0.019&0.098\\
		\hhline{|~|-|-|-|-|-|-|-|-|-|-|-|-|-|-|-|-|-|}
		
		&{SSP}&
		0.697&0.580&0.253&0.000&
		\textbf{0.293}&1.000&0.949&1.000&
		0.000&0.000&0.000&0.000&
		0.481&0.007&0.069&0.050\\
		\hhline{|~|-|-|-|-|-|-|-|-|-|-|-|-|-|-|-|-|-|}
		
		&{SST}&
		0.588&0.696&\cellcolor{LightBlue2}\textbf{0.918}&0.167&
		\textbf{0.400}&0.777&0.624&0.875&
		0.000&0.000&0.000&0.000&
		0.419&0.005&0.333&0.144\\
		\hhline{|~|-|-|-|-|-|-|-|-|-|-|-|-|-|-|-|-|-|}
		
		&{MCGS (Ours)}&
		\cellcolor{LightBlue2}\textbf{1.000}&\cellcolor{LightBlue2}\textbf{1.003}&\cellcolor{LightBlue2}\textbf{1.013}&\cellcolor{LightBlue2}\textbf{1.837}&
		\cellcolor{LightBlue2}\textbf{0.000}&\cellcolor{LightBlue2}\textbf{0.000}&\cellcolor{LightBlue2}\textbf{0.206}&\cellcolor{LightBlue2}\textbf{0.341}&
		\cellcolor{LightBlue2}\textbf{0.910}&\cellcolor{LightBlue2}\textbf{0.803}&\cellcolor{LightBlue2}0.497&\cellcolor{LightBlue2}\textbf{0.870}&
		\cellcolor{LightBlue2}0.170&0.003&\cellcolor{LightBlue2}1.000&\cellcolor{LightBlue2}0.209\\
		\hline

	\end{tabular}
	\vspace*{-0.5cm}
\end{table*}
%\vspace{-2cm}
%sec5

\section{{New Algorithm Proposal}}
\label{sec:sec5}
The results of the experimental study confirm that the existing algorithms can hardly preserve minority structures. In this section, we introduce a new algorithm called MCGS.

%5.1
\subsection{Definitions and Notations}
\label{sec:sec5.1}
A $graph$ is notated with ${G = (V, E)}$, where $V = $\{$v_1, v_2, ... , v_n$\}$ $ represents nodes, $E = $\{$e_1, e_2, ... , e_m $\}$ $ represents edges, and an edge $e = (v_i, v_j)$ connects nodes ${v_i}$ and ${v_j}$. This work focuses on scale-free graphs \cite{E10} and stipulates that graphs are simple, unattributed, undirected, and connected to simplify the representations.

\par{A \textit{graph sample} is notated with $G_s = (V_s, E_s)$, where $V_s$ is a subset of nodes ($V_s\subset V$) and $E_s = (V_s\times V_s) \cap E$. Considering a node-based sampling strategy, a \textit{sampling rate} is defined as $\Phi = \mid V_s\mid /\mid V\mid$, where $\mid V_s\mid$ is the cardinality of $V_s$ and $\mid V\mid$ is the cardinality of $V$. $E_s$ is obtained from the induced subgraph of $G$ based on $V_s$.}

\par{We use $\Omega=$\{$P,S,R,T$\}$ $ to represent the four types of minority structures in $G$, where $P$ represents the set of all super pivots, notated by $P=$\{$p_1,…,p_l$\}$ $, whereas $S$, $R$, and $T$ represent the sets of all huge stars, rims, and ties, respectively. The four types are defined as follows: }

\par{A super pivot is a node whose one-step neighbor nodes have at least one interconnection, with its degree within the global top 5$\%$ as represented by $\mu$. Super pivot is notated as $\forall p_i\in P$: (1) $\exists!v\in V_{p_i}$: $\mid\Gamma(v)\mid\geq\mu$, where $V_{p_i}$ is the node set of $p_i$, $\Gamma(v)$ is the set of one-step neighbors of $v$, $\mid\Gamma(v)\mid$ is the cardinality of $\Gamma(v)$; and (2) $\exists x\in\Gamma(v)$: $\mid\Gamma(x)\cap\Gamma(v)\mid\geq1$.}

\par{A huge star is a node whose one-step neighbor nodes are not connected to one another, with its degree above the global mean as represented by $\epsilon$. Huge star is notated as $\forall s_i\in S$: (1) $\exists!v\in V_{s_i }$: $\mid\Gamma(v)\mid\geq\epsilon$; (2) $\forall x\in \Gamma(v)$: $\Gamma(x)\cap\Gamma(v)=\emptyset$.}

\par{A rim is a node appearing at community margins and forming a parachute-like visual shape with its one-step neighbors or a sequence of nodes forming a chain-like visual shape, notated as $\forall r_i\in R$: (1) $V_{r_i}\subset V_{c_i}$, where $C=$\{$c_1,…,c_n$\}$ $ is the set of all disjoint communities created from $G$, and $V_{c_i}$ is the node set of $c_i$; (2) $\exists v\in V_{r_i}$: $v\in V_{cut}\wedge\mid\Gamma(v)\mid\geq2$, where $V_{cut}$ is the set of cut points of $G$; and (3) $\forall x\in V_{r_i}-V_{cut}$: $\mid\Gamma(x)\mid=1\wedge\Gamma(x)\subset V_{r_i}$.}

\par{A tie is a sequence of nodes that bridge any two communities and form a chain-like visual shape, notated as $\forall t_i\in T$: (1) $V_{t_i}\subset V_{cut}$; (2) $\exists!u,v\in V_{t_i}$: $u\in V_{c_j}\wedge v\in V_{c_k}\wedge j\neq k$; and (3) $\forall x\in V_{t_i}-$\{$u,v$\}: $\mid\Gamma(x)\mid=2\wedge\Gamma(x)\subset V_{t_i}$.}

%5.2
\subsection{Design Considerations}
On the basis of the experience and results of the experimental study, we formulate six key points to be considered in the algorithm design.

\par{\textbf{C1:} Identifying minority structures. Effectively preserving minority structures through global random sampling is difficult because minority structures only involve a small proportion of nodes. A feasible way is to identify and maintain them in advance.}

\par{\textbf{C2:} Preserving important minority structures. Prioritizing important minority structures is necessary for two reasons, that is, reserving space in a sample for subsequent majority structure sampling and minimizing the influence of random sampling on minority structure preservation.}

\par{\textbf{C3:} Preserving neighbors of minority structures. Simply selecting the self-nodes of minority structures is insufficient because they are no longer minority structures if losing neighbors.}

\par{\textbf{C4:} Balancing minority and majority structures. The absence of majority structures invalidates the sample because minority structures coexist with majority structures.}

\par{\textbf{C5:} Suppressing the generation of new minority structures. Existing algorithms produce new minority structures, possibly leading to a misjudgment on the graph by sample analysis.}

\par{\textbf{C6:} Improving robustness and scalability. For robustness, the influence of parameter settings and graph data sets on sampling should be minimized. For scalability, sampling should be completed within an acceptable time on large-scale graphs.}

%5.3
\subsection{Algorithm Pipeline}
\begin{figure*}[htbp]
	\centering
	\vspace{-0.2cm}  %调整图片与上文的垂直距离	
	\setlength{\abovecaptionskip}{1pt}   %调整图片标题与图距离	
	\setlength{\belowcaptionskip}{-0.5cm}   %调整图片标题与下文距离
	\includegraphics[width=\linewidth]{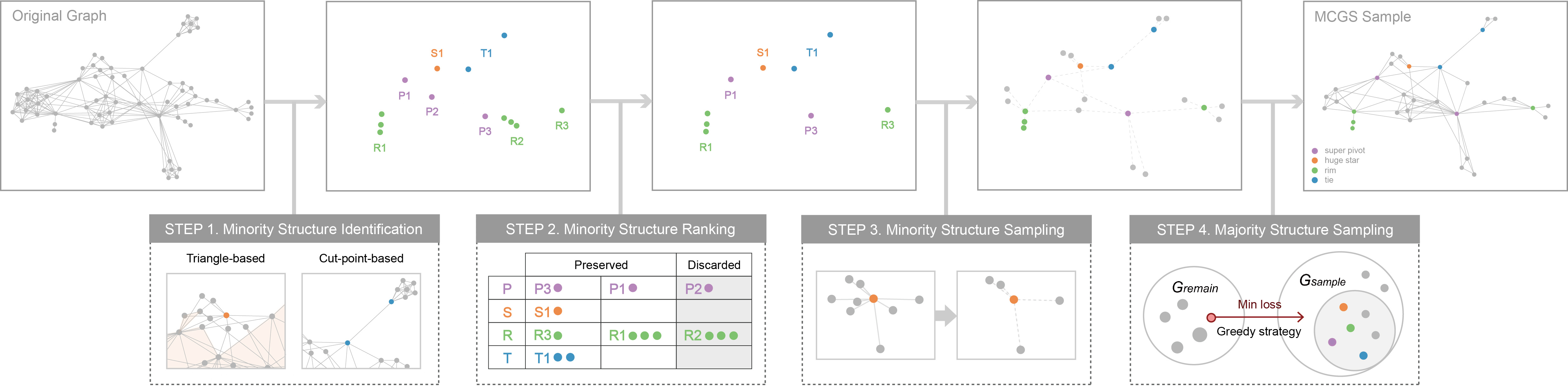}
	\caption{Four-step pipeline of MCGS: (1) identifying minority structures by using a triangle-based algorithm and a cut-point-based algorithm; (2) ranking minority structures based on importance assessment criteria; (3) preserving important minority structures and randomly selecting their neighbors into the sample according to the sampling rate (50$\%$ in this case); (4) selecting other nodes into the sample based on a greedy optimization strategy and outputting the induced subgraph as the completed sample.}
	\label{fig:figure2}
\end{figure*}

The proposed MCGS algorithm consists of four steps, as shown in \autoref{fig:figure2}.

\par{\textbf{STEP1.} Minority structure identification. Given a graph $G$, a sample $G_s$, and a sampling rate $\Phi$, we identify minority structures in $G$ by using two newly designed algorithms (Section \ref{sec:sec5.4.1}). This step outputs the four sets of minority structures $\{P,S,R,T\}$ that contain all super pivots, huge stars, rims, and ties in $G$, respectively.}

\par{\textbf{STEP2.} Minority structure ranking. We initially rank the four sets of minority structures separately in descending order of importance by using our proposed importance assessment criteria (Section \ref{sec:sec5.4.2}) and quick sort. Then, we select the most important ones in each of the four sets based on $\Phi⁄\alpha$, where $\alpha$ is a constant that controls the quantity of preserved minority structures and is set to 1 by default. {Specifically, a smaller $\alpha$ indicates more minority structures to be preserved.} For example, given a graph with 3 ordered rims and $\Phi=$ 50$\%$, we pick the top two $\lceil 3\times(0.5/1) \rceil$ important ones. This step outputs $\{P_{im},S_{im},R_{im},T_{im}\}$.}

\par{\textbf{STEP3.} Minority structure sampling. We directly put all nodes in $\{P_{im},S_{im},R_{im},T_{im}\}$ into $G_s$ and then randomly select a proportion of their one-step neighbor nodes into $G_s$ by using an improved RAS sampling (Section \ref{sec:sec5.4.3}). This step outputs an incomplete $G_s$ that contain nodes of all important minority structures and parts of their neighbors.}

\par{\textbf{STEP4.} Majority structure sampling. We propose a greedy strategy to select the nodes in $G$ to $G_s$ to maximize the similarity between $G_s$ and $G$. After reaching the sampling rate, we preserve all edges in the induced subgraph from $G$ based on $G_s$ to suppress the generation of new minority structures (see Section \ref{sec:sec5.4.4}). This step outputs the completed $G_s$.}

\par{We also provide an optional additional step, that is, unbalanced graph processing. If $G$ is examined as an unbalanced graph, then $G$ will be divided into several subgraphs, and the above sampling process will be conducted on each of them. This step can reduce the influence of unbalanced graphs on the minority structure preservation (C6). This step is optional because a majority of graphs are balanced. We use a method based on the gradient boosting decision tree \cite{E1,add1} and a multilevel partitioning method \cite{E2} in this step. Supporting information for this step is provided in the supplementary material.}

%5.4
\subsection{{Algorithm Components}}
%5.4.1
\subsubsection{Minority Structure Identification}
\label{sec:sec5.4.1}
STEP 1 is to identify minority structures in $G$ (C1). We propose two fast identification algorithms because straightforward methods are generally time-consuming (C6).

\par{A straightforward method of identifying pivots and stars is to seek out high degree nodes and check whether edges exist between its neighbors. If no edge exists, then it is a star; otherwise, it is a pivot. This method is time-consuming with a time complexity of $O(n^2)$, where $n$ is the number of nodes in $G$. We design a triangle-based algorithm inspired by two ideas: a high degree node is either a pivot or a star, facilitating a simultaneous detection of pivot and star; a node whose relationships with any two of its neighbors form a triangle cannot be a star, accelerating the process.}

\par
\begin{figure}[htbp]
	\centering
	\vspace{-0.03cm}  %调整图片与上文的垂直距离	
	\setlength{\abovecaptionskip}{-0.01cm}   %调整图片标题与图距离	
	\setlength{\belowcaptionskip}{-0.2cm}   %调整图片标题与下文距离
	\includegraphics[width=\linewidth]{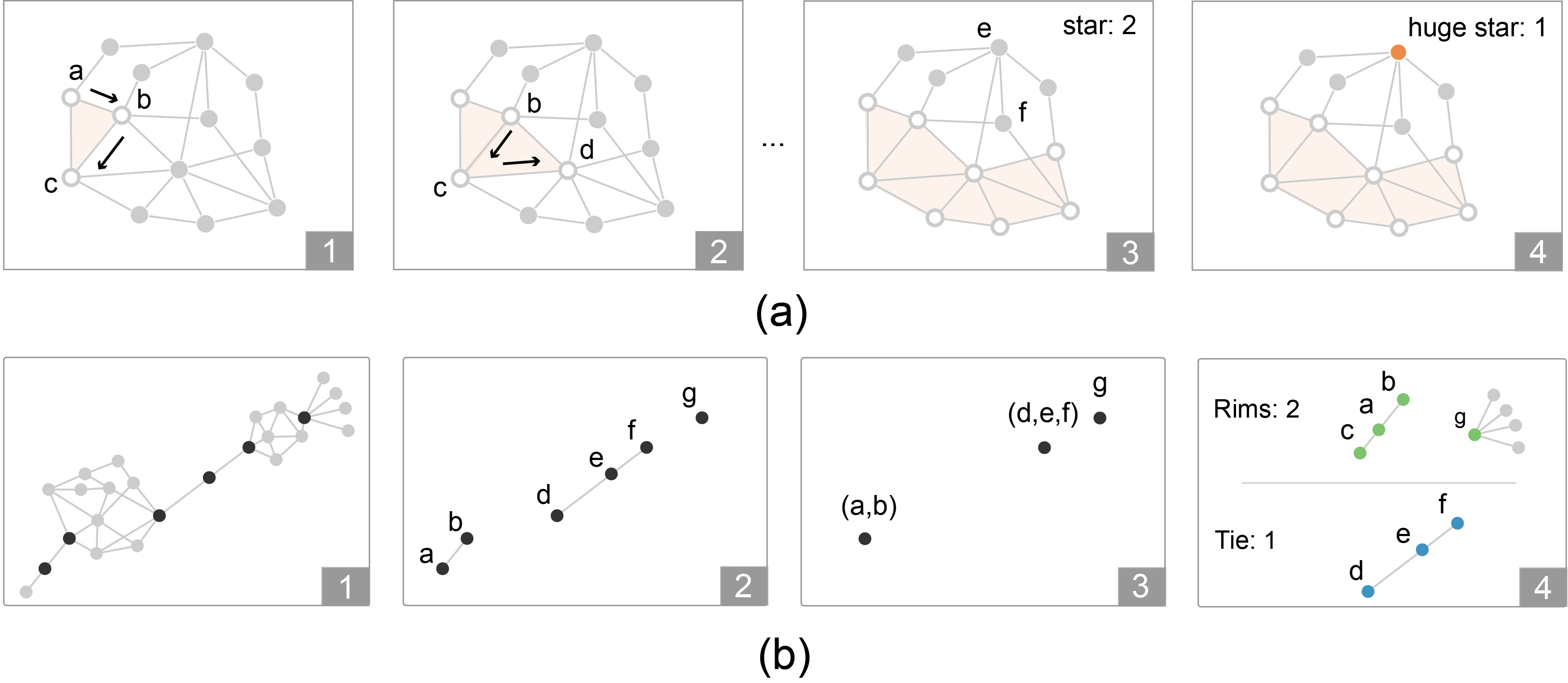}
	\caption{Illustration of triangle-based super pivot and huge star identification (a) and cut-point-based rim and tie identification (b).}
	\label{fig:figure3}
\end{figure}

{This algorithm consists of five steps. (1) Given a graph, a DF traversal starts from any node. (2) For a visiting node, we check whether it forms a triangle with its predecessor and the node preceding the predecessor. If so, then we mark all the three nodes, such as the nodes marked with hollow dots in \autoref{fig:figure3}(a-1) and 3(a-2). (3) After the traversal, unmarked high degree nodes are identified as stars, such as the nodes e and f in \autoref{fig:figure3}(a-3). (4) We identify high degree nodes in $G$ but not in the set of unmarked nodes as pivots. (5) We extract pivots with degrees within the global top 5$\%$ as super pivots and stars with degrees above the global mean as huge stars, such as the huge star e in \autoref{fig:figure3}(a-4). The time consumption of this algorithm mainly arises from the DF traversal with a complexity of $O(n + m)$, where $n$ and $m$ are the numbers of nodes and edges in $G$, respectively.}

\par{{A straightforward method to identify rims and ties is to use community information. However, community detection methods are time-consuming and complicated in parameter tuning. For example, it is difficult to determine the number of communities, which directly influences the identification of rims and ties (C6). We propose to use cut-point information because both a rim and tie have at least one cut point. A cut point is a node whose removal will cause the relevant connected subgraph to be disconnected. \autoref{fig:figure3}(b-1) highlights all cut points in a graph.}}

\par{The cut-point-based algorithm has four steps. (1) Given a graph, we obtain all cut points by DF traversal \cite{E3} to generate an induced subgraph denoted as $G_{cut} = (V_{cut}, E_{cut})$. (2) We merge each connected component of $G_{cut}$ into a hyper node, as shown in \autoref{fig:figure3}(b-2) and 3(b-3). (3) We identify a hyper node that contains only one original node as a parachute-like rim. (4) A hyper node point that contains multiple original nodes is a chain structure. If any end node of the chain has one and only one neighbor with degree 1 in $G$, then all nodes in the chain together with the neighbor are identified as a chain-like rim; otherwise, all nodes in the chain are regarded as a tie. The time complexity of this algorithm is $O(n + m)$. In addition, large parachute-like rims may be identified as super pivots or huge stars. This case rarely appears because the degree of rim is generally not high. In our work, such large rims are counted as not only rims but also super pivots or huge stars.}

%5.4.2
\subsubsection{Minority Structure Ranking}
\label{sec:sec5.4.2}
Essential to STEP 2 is to determine importance assessment criteria for each minority structure type (C2). The results of the pilot user study reflect that the degree or size of a minority structure is directly related to its visual importance. We adopt this empirical result, which is simple and efficient. For a super pivot or a huge star, we stipulate its importance proportional to its degree. The importance of a parachute-like rim is proportional to the number of its neighbors with a degree of 1. For chain-like rims, a long chain is important. For a tie, we consider two factors, namely, the chain length and number of neighbors connecting to both ends of the chain.

%5.4.3 
\subsubsection{Minority Structure Sampling}
\label{sec:sec5.4.3}
STEP 3 is to preserve the neighborhood structures of important minority structures (C3). RAS provides an efficient and effective way that directly maintains all one-step neighbors \cite{B43}. However, this way may include extensive nodes in a sample, thereby causing an early reaching of the sampling rate (C4). We propose to add a specific condition to RAS sampling. The amount of preserved neighbors for a minority structure should satisfy the condition $\mid\Gamma_s(ms_i)\mid=\mid\Gamma(ms_i)\mid* \Phi/\beta$, where $ms_i$ denotes the key nodes of the minority structure $i$, $\Gamma(ms_i)$ is the set of neighbors of $ms_i$, $\Gamma_s(ms_i)$ is the set of preserved neighbors of $ms_i$. For a super pivot, huge star, or parachute-like rim, $ms_i$ includes only one node; whereas for a chain-like rim or tie, $ms_i$ includes the end nodes of the chain.  $\beta$ is a constant that controls the quantity of preserved neighbors; it is set to 2 by default to reserve half of the sample space for majority structure preservation (STEP4). {Using $\alpha$ and $\beta$ altogether can tune the preserving ratio of minority structures versus majority structures.} 

%5.4.4
\subsubsection{Majority Structure Sampling}
\label{sec:sec5.4.4}
An incomplete sample $G_s$ that contains all important minority structures and parts of their neighbor nodes is obtained. STEP 4 aims to further add nodes and edges from $G$ to $G_s$, making the completed sample $G_s$ as similar to $G$ as possible (C4). This situation can be described as an optimization problem. Given a $G = (V, E)$, an incomplete sample $G_s = (V_s, E_s)$, and a sampling rate $\Phi$, an optimal node set $V_{op}$ and an edge set $E_{op}$ are added to $G_s$ to ensure that the completed sample $G_s$ can effectively represent $G$, where $V_{op}\subset$\{$V-V_s$\}$ $, $E_{op}\subset$\{$E-E_s$\}$ $, and $\mid V_{op}\mid =\mid V\mid*\Phi - \mid V_s\mid$. The objective function is as follows:
\begin{equation}
\setlength{\abovedisplayskip}{4.5pt}
\setlength{\belowdisplayskip}{0pt}
G^*_s \leftarrow \arg\mathop{\min}_{G_s \subset G} Deviance(G, G_s),
\nonumber
\end{equation}

\begin{equation}
\setlength{\abovedisplayskip}{0pt}
\setlength{\belowdisplayskip}{4.5pt}
\mathrm{ s.t.}\mid V_s\mid = \mid V\mid \times\Phi,
\nonumber
\end{equation}

\par{This problem is NP-hard. Classical optimization algorithms, such as genetic algorithm \cite{E4} and simulated annealing \cite{E5}, are candidates for problem solving but commonly have high computational consumptions and complicated parameter settings (C6). We adopt a greedy strategy that has no parameters, a high speed, and desired effects. The strategy consists of four steps. (1) For each node in $\left\{V-V_s\right\}$, we suppose to add it to $V_s$ and then calculate the deviance of $G$ and $G_s$ (the induced subgraph of $G$ based on $V_s$), called $loss$. (2) We find the minimum loss obtained in (1) and add the corresponding node into $V_s$ formally. (3) We repeat (1) and (2) until $\mid V_s\mid$ reaches $\mid V\mid\times\Phi$. (4) We output the induced subgraph of $G$ based on $V_s$ as the completed sample $G_s$. The complexity of the strategy is $O(\mid V-V_s\mid\times(n+m))$.}

\par{{The induction step is important. It can repair the neighborhood structures of sampled nodes and make the incomplete structures close to that of the original graph \cite{B24}. As a result, the neighborhoods of potential new stars, rims and ties can be repaired, thereby effectively suppressing the generation of new minority structures (C5)}. We are inspired by RDN, RPN, and TIES that obtained the distinguished performance on the MSGR indicator in the experimental study because they adopted an induction step.}

\par{The definition of the loss function is critical in the greedy strategy. The ultimate goal of majority structure sampling is to make $G_s$ as similar to $G$ as possible. Such similarities are commonly measured by the metrics mentioned in Section \ref{sec:sec2.3}. We reference three popular metrics, namely, DD \cite{A4}, NCC \cite{F7}, and JI \cite{E9}, to propose three objectives as follows.}

\par{(1) We use the mean square error (MSE) of degrees between $G_s$ and $G$ to depict the similarity of DD, notated as:
\begin{equation}
\setlength{\abovedisplayskip}{4.5pt}
\setlength{\belowdisplayskip}{4.5pt}
MSE = \frac{1}{\mid V_s \mid}\sum_{i=1}^{\mid V_s\mid}(\mid\Gamma(G, v_i)\mid-\mid\Gamma(G_s, v_i)\mid)^2,
\nonumber
\end{equation}
where $\Gamma(G, v_i)$ is the set of neighbors of $v_i$ in $G$}

\par{(2) We use NCC($G_s$) to represent the number of connected components of $G_s$, which can measure the similarity of connectivity between $G_s$ and $G$ because $G$ is supposed to be connected in this work, thus NCC($G$)=1, notated as:
\begin{equation}
\setlength{\abovedisplayskip}{4.5pt}
\setlength{\belowdisplayskip}{4.5pt}
NCC(G_s)=\mid V_s\mid-\sum\nolimits_{i\in\left[\ \mid V_s\mid \right]}\ \mbox{\uppercase\expandafter{\romannumeral2}}\left[\ \sigma_i(L)>0\right],
\nonumber
\end{equation}
where $L$ is the Laplacian matrix of $G$, $L\in R^{n\times n}$, $\sigma_i (L)'s$ are the singular values of $L$.}

\par{(3) We use the Jaccard Index (JI) to measure the structural similarity between $G_s$ and $G$, notated as:}
\begin{equation}
\setlength{\abovedisplayskip}{4.5pt}
\setlength{\belowdisplayskip}{4.5pt}
JI(G,G_s)=\frac{1}{\mid V_s\mid}\sum_{v_i\in V_s}\frac{\mid\Gamma(G,v_i)\cap\Gamma(G_s,v_i)\mid}{\mid\Gamma(G,v_i)\cup\Gamma(G_s,v_i)\mid}.
\nonumber
\end{equation}

\par{As a result, our loss function is defined as:
\begin{equation}
\setlength{\abovedisplayskip}{4.5pt}
\setlength{\belowdisplayskip}{4.5pt}
loss=\omega_1 * Scaler(MSE)+\omega_2*Scaler(NCC(G_s))+\omega_3*Scaler(JI(G,G_s)),
\nonumber
\end{equation}
where $\omega_i$ is a weight coefficient,$\omega_i\in\left[\ 0,1\right]$, $\sum_{i=1}^{3}\omega_i=1$; and $Scaler$ is a normalization processing to reduce the influence of the magnitude difference among the three objective functions.}

\par{The computation of the loss function could be accelerated (C6). We could only consider the incremental information of adding a new node into $G_s$ each time when calculating MSE. We could use the union-find algorithm \cite{E6} to immediately obtain the number of disjoint sets of a graph for NCC. We could adjust the three weight coefficients on demand to involve only one or two objectives in computation. We set them with [1:0:0] by default. Empirically, no single sampling method can simultaneously fulfill the optimal effects on the three objectives \cite{E7,E8}.}

%sec6
\section{Evaluation}
We evaluated the proposed MCGS algorithm through an objective performance analysis, a subjective assessment, and case studies.
%6.1
\subsection{Objective Performance Analysis}
The performance analysis had three experiments with different indicators. The data, reference algorithms, execution conditions, result processing, and apparatus of the experiments were consistent with those of the experiment introduced in Section \ref{sec:sec4}.
%6.1.1
\subsubsection{Minority Structure Preservation Performance}
\label{sec:sec6.1.1}
Indicators in this experiment were MSPR, MSGR, and MIP, which can evaluate the performance of minority structure preservation (Section \ref{sec:sec4.3}). The results of MCGS are shown in the last row of \autoref{tab:table2}. We conducted 12 groups of significance tests (3 indicators × 4 minority structure types) for MCGS and the 20 references algorithms. We initially used Shapiro-Wilk tests for each group to examine the normality of the experimental results of each algorithm on the 10 graphs and four sampling rates. The examination results did not follow the normal distribution. Then, we used a non-parametric Friedman test for each group. Significant differences  ($p$ $<$ 0.05) were found in all the 12 groups. Finally, we used a Dunn–Bonferroni test to identify the winners in each group.

\par{The results show that MCGS won 12 times in the significance tests and performed best 11 times in terms of indicator medians. MCGS underperformed RMSC on the MSGR medians of ties because of the superiority of RMSC in suppressing new ties. MCGS did not obtain a “good” MIP median on rims because MCGS could not completely avoid the generation of new minority structures, and new rims were most apt to be produced among the four types. In summary, MCGS overall performed best among the 20 references. It could effectively preserve the four types of minority structures, suppress the generation of new minority structures, and prevent the loss of important minority structures.}

%6.1.2
\subsubsection{Majority Structure Preservation Performance}
\label{sec:sec6.1.2}
Indicators in this experiment were Kolmogorov–Smirnov distance (KSD) \cite{F1}, skew divergence distance (SDD) \cite{F2}, reciprocal of NCC (RCC) \cite{G32}, and JI \cite{E9}. They are commonly used in graph sampling evaluations (Section \ref{sec:sec2.3}). KSD and SDD measure the difference of degree distributions between a graph and sample, RCC measures the connectivity differences before and after sampling, and JI measures the similarity of graphs. The four indicators range from 0 to 1. Small values for KSD and SDD and large values for RCC and JI are good. The experimental results are shown in the four rightmost columns of \autoref{tab:table2}. Notably, we tested significant differences but did not provide empirical “good” thresholds in the results.

\par{Significant differences were found in the four indicators among the 21 algorithms. MCGS became the winners of KSD, RCC, and JI, indicating that MCGS did not pursue minority structure preservation at the expense of majority structure preservation. For SDD, MCGS obtained a satisfying median. Thus, the performance of MCGS in preserving majority structures was fairly satisfactory.}

%6.1.3
\subsubsection{Time Performance}
The indicator in this experiment was the mean time consumption of 20 samplings (4 seed types × 5 runs) performed by an algorithm on a graph under a sampling rate. We tested 21 algorithms, 10 graphs, and four sampling rates. Due to page limit, we only show the results of MCGS and seven reference algorithms under a sampling rate of 30$\%$ on four graphs (\autoref{tab:table3}). More results are provided in the supplementary material.

\begin{table}[!htbp]
	%\tiny
	\scriptsize
	%fontsize{5pt}{1.5}
	\renewcommand\tabcolsep{4pt}  %% 调整表格列间的宽度
	\renewcommand\arraystretch{1.2} %设置表格行间距
	\centering
	\vspace{-0.1cm}  %调整图片与上文的垂直距离
	\caption{Time consumptions of MCGS and seven reference algorithms on four graph data sets.}
	\label{tab:table3}
	\begin{tabular}{|c|c|c|c|c|}%一个c表示有一列，格式为居中显示(center)
		\hline  % 在表格最上方绘制横线
		\rule{0pt}{5pt}\multirow{4}{*}{\textbf{Algorithm}}&
		\multicolumn{4}{c|}{\textbf{Time Consumptions on Graph Data Sets (Sec.)}}\\
		\cline{2-5}
		\vspace{-2.5pt}
		\rule{0pt}{5pt}&{\textbf{Facebook1912}}&{\textbf{Facebook107}}& {\textbf{AS-733}}&{\textbf{PGP}}\\
		\vspace{-2.5pt}
		&{\textit{node:747}}&{\textit{node:1034}}&{\textit{node:6474}}&{\textit{node:10680}}\\
		%\vspace{-1pt}
		&{\textit{edge:30025}}&{\textit{edge:26749}}&{\textit{edge:13895}}&{\textit{edge:24316}}\\

		\hline
			
		\rule{0pt}{5pt}{{TIES}}&0.0066&0.0071&0.0075&0.0146\\
		\hline		
		\rule{0pt}{5pt}{{RMSC}}&0.0006&0.0012&0.0058&0.0340\\
		\hline	
		\rule{0pt}{5pt}{{FF}}&0.0021&0.0030&0.0121&0.0355\\
		\hline	
		\rule{0pt}{5pt}{{RDN}}&0.0019&0.0033&1.5038&0.4520\\
		\hline	
		\rule{0pt}{5pt}{{DPL}}&1.0846&1.5198&3.4403&25.3200\\
		\hline
		\rule{0pt}{5pt}{{MCGS (Ours)}}&0.7712&0.9078&9.4609&34.7743\\
		\hline
		\rule{0pt}{5pt}{{DLAS}}&17.9142&14.6273&10.7340&21.3550\\
		\hline
		\rule{0pt}{5pt}{{SST}}&0.4130&0.8868&28.7397&83.3247\\
		\hline
	\end{tabular}
	\vspace{-0.2cm}
\end{table}

\par{{We found that the eight algorithms can be divided into two groups. The first group included TIES, RMSC, FF, and RDN. Their time consumptions were lower than those of algorithms in the second group, which included DPL, MCGS, DLAS, and SST. The main reason was that the algorithms of the second group commonly had additional computation steps during random sampling. For example, DPL detected communities and SST generated spanning trees. Such steps in MCGS (i.e., minority structure identification, importance ranking, and loss function computation) were not very time-consuming. Therefore, MCGS presented relatively good time consumptions in the second group. Moreover, MCGS was very fast on the Facebook1912 and Facebook107 due to its insensitivity to the scale of edges. In summary, the time performance of MCGS was at the low-medium level in the experiment. The time performance can be further improved by adopting parallel computations or simplifying the greedy strategy by selecting the optimal node out of random nodes rather than all remained nodes.}}

%6.2
\subsection{Subjective Assessment}
\begin{figure}[!h]
	\centering
	\vspace{-0.2cm}  %调整图片与上文的垂直距离	
	\setlength{\abovecaptionskip}{5pt}   %调整图片标题与图距离	
	\setlength{\belowcaptionskip}{-0.4cm}   %调整图片标题与下文距离
	\includegraphics[width=\linewidth]{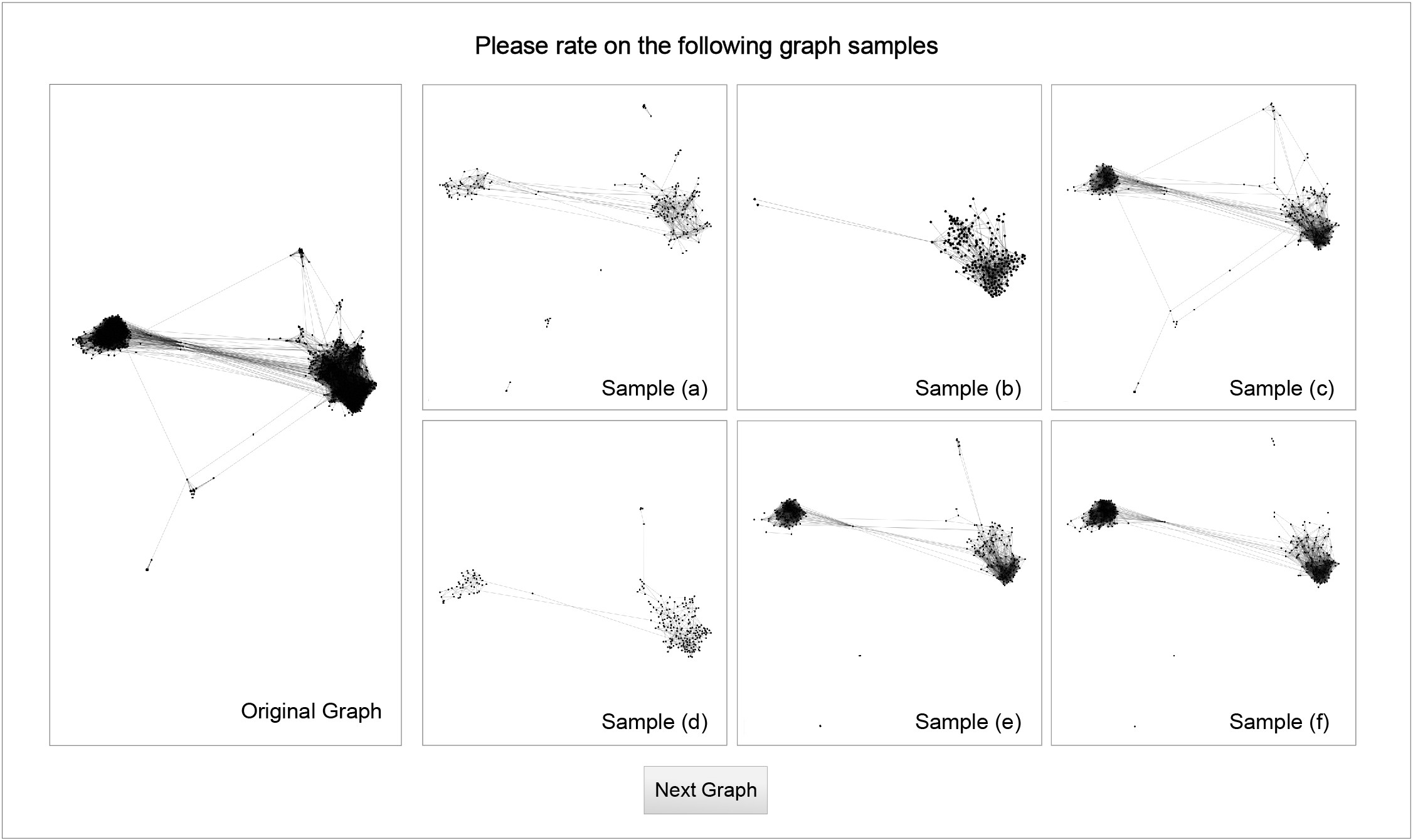}
	\caption{Interface for subjective assessment experiment with the Facebook1684 data set (775 nodes and 14,006 edges). The leftmost graph is the original graph. The six randomly arranged samples are generated by SST (a), FF (b), MCGS (c), RW (d), TIES (e), and RDN (f) with a sampling rate of 30$\%$.}
	\label{fig:figure4}
\end{figure}

We recruited the 20 participants in the pilot user study again to conduct a subjective assessment experiment. They were asked to assess similarities between a graph and samples by perceiving node-link diagrams \cite{F6} and rating on six metrics. Three of the metrics were similarities of the overall shape, community, and connectivity for assessing majority structure preservation. The other three metrics were similarities of high degree, margin, and boundary structures related to the preservation of minority structures. We selected eight popular graphs from the 34 graphs used in the user study. We selected the proposed MCGS and five reference algorithms, namely, RDN, TIES, FF, RW, and SST, most of which performed relatively well in the previous experiments. We used the same high degree nodes as initial seeds. The sampling rate was set at 30$\%$, which is empirically suitable for visual perception \cite{B23,B48}. A graph and six randomly arranged samples were presented at a time (\autoref{fig:figure4}). The participants rated on each sample from the six metrics with a five-point Likert scale ranging from 1 (the lowest similarity) to 5 (the highest similarity).

\par{The results of the top three algorithms on each metric in terms of average rating are as follows. (1) Overall shape: MCGS ($\mu$ = 4.3), TIES ($\mu$ = 3.7), and RDN ($\mu$ = 3.5); (2) Community: MCGS ($\mu$ = 4.3), TIES ($\mu$ = 3.8), and RDN ($\mu$ = 3.7); (3) Connectivity: MCGS ($\mu$ = 4.3), TIES ($\mu$ = 3.6), and RDN ($\mu$ = 3.2); (4) High degree structure: MCGS ($\mu$ = 4.1), TIES ($\mu$ = 3.7), and RDN ($\mu$ = 3.6); (5) Margin structure: MCGS ($\mu$ = 4.0), TIES ($\mu$ = 3.0), and SST ($\mu$ = 2.9); (6) Boundary structure: MCGS ($\mu$ = 4.1), TIES ($\mu$ = 3.1), and SST ($\mu$ = 2.9). MCGS obtained the highest average rating six times, followed by TIES and RDN with relatively high average ratings. The results reflected that the minority structure preservation affected perceived similarities to a certain extent and the MCGS samples achieved considerable perceived similarities. In the interview, we focused on similarity judgment principles. Most of the participants stated that they initially observed the overall shape and connectivity and then considered visually prominent minority structures.}

%6.3
\subsection{Case Studies}
We used three popular graph data sets to demonstrate the features of MCGS. The reference algorithms, initial seeds, sampling rate, and layout method were consistent with the subjective assessment experiment. In node-link diagrams, the relative locations of nodes in a sample are consistent with those of the corresponding nodes in the original graph. Additional cases are provided in the supplementary material.

%6.3.1
\subsubsection{AS-733 Graph Data Set}
\begin{figure*}[htbp]
	\centering
	\vspace{-0.3cm}  %调整图片与上文的垂直距离	
	\setlength{\abovecaptionskip}{-0.03cm}   %调整图片标题与图距离	
	\setlength{\belowcaptionskip}{-0.3cm}   %调整图片标题与下文距离
	\includegraphics[width=\linewidth]{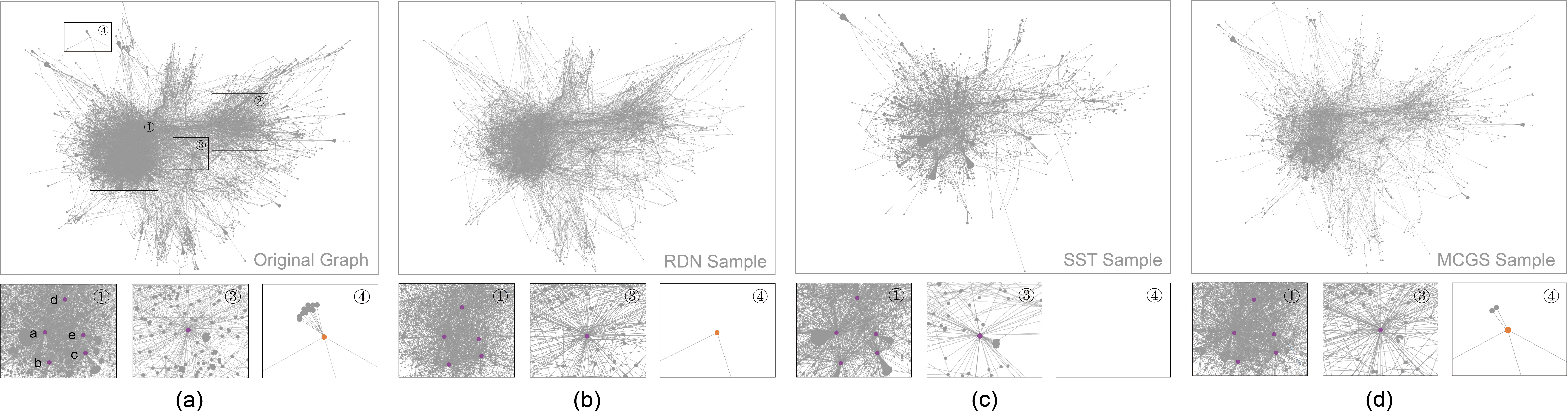}
	\caption{Case analysis of the AS-733 graph data set (a) and three samples generated by RDN (b), SST (c), and MCGS (d) with a sampling rate of 30$\%$.}
	\label{fig:figure5}
\end{figure*}

The AS-733 graph data set \cite{B27} is an autonomous systems network on the Internet with 6,474 nodes and 13,895 edges. The original graph and samples obtained by RDN, SST, MCGS are shown in \autoref{fig:figure5}.

\begin{figure*}[!h]
	\centering
	\vspace{0.1cm}  %调整图片与上文的垂直距离	
	\setlength{\abovecaptionskip}{1pt}   %调整图片标题与图距离	
	\setlength{\belowcaptionskip}{-0.5cm}   %调整图片标题与下文距离
	\includegraphics[width=\linewidth]{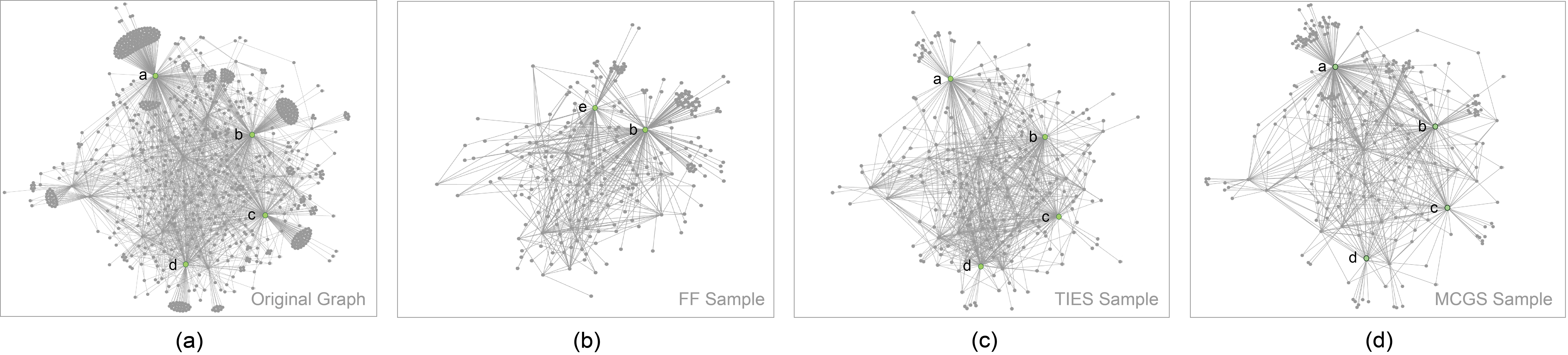}
	\caption{Visual illustration of the Cpan graph data set (a) and three samples generated by FF (b), TIES (c), and MCGS (d) with a sampling rate of 30$\%$.}
	\label{fig:figure6}
\end{figure*}

\par{We marked four SOIs popular in the pilot user study, as shown in \autoref{fig:figure5}(a). SOI-1 and SOI-2 were the first and second largest communities in the graph, respectively. SOI-3 was a visually prominent super pivot in a relatively sparse area. SOI-4 was a huge star far way the two communities. In SOI-1, the top five important super pivots were ranked and named as (a [degree = 1,460] $>$ b [degree = 752] $>$ c [degree = 693] $>$ d [degree = 401] $>$ e [degree = 378]) in a descending order of degree.}

\par{For the RDN sample in \autoref{fig:figure5}(b), the overall shape and density distribution of the original graph were considerably preserved. The five super pivots in SOI-1 were maintained, but only the top two super pivots (a [586] $>$ b [367] $>$ d [350] $>$ c [205] $>$ the other [145]) were consistent with those in the original graph in order. The super pivot in SOI-3 was well preserved. The huge star in SOI-4 was retained but lost many neighbors.}

\par{The SST sample in \autoref{fig:figure5}(c) was of low similarity with the original graph in the overall shape and density distribution. The second largest community in SOI-2 and the huge star in SOI-4 disappeared. The five super pivots in SOI-1 were maintained. The top three important super pivots (a [499] $>$ b [228] $>$ c [196] $>$ the other [156] $>$ d [118]) were preserved in order; however, most of their preserved neighbors were the nodes with the degree of 1 and many inter-connections in the community lost. This situation was also reflected in SOI-3.}

\par{For the MCGS sample in \autoref{fig:figure5}(d), the overall shape and the density distribution were well preserved. The five super pivots in SOI-1 were maintained, and the top four important ones (a [589] $>$ b [338] $>$ c [301] $>$ d [196] $>$ the other [179]) were consistent with those in the original graph in order. The super pivot in SOI-3 and the huge star in SOI-4 were well preserved.}

\par{Among the three algorithms, RDN performed best on majority structure preservation, followed by MCGS. MCGS performed best on preserving super pivots and huge stars, especially for the maintenance of the importance order of super pivots. SST only performed well in preserving one-degree neighbors of super pivots.}

%6.3.2
\subsubsection{Cpan Graph Data Set}
The Cpan data set is a collaboration network with 839 nodes and 2,127 edges \cite{F4}. It depicts the relationships between the developers using the same Perl modules. The original graph and samples obtained by FF, TIES, and MCGS are shown in \autoref{fig:figure6}. This case focused on the preservation of parachute-like rims at marginal areas. The top four important rims in the original graph were marked as a $>$ b $>$ c $>$ d in descending order in \autoref{fig:figure6}(a). The FF sample in \autoref{fig:figure6}(b) presented an overall shape dissimilar to the original graph, and only one of the four important rims and another rim were maintained. The TIES sample in \autoref{fig:figure6}(c) presented an overall shape greatly similar to the original graph, and the four important rims were preserved with a changed importance order (a $>$ d $>$ b $=$ c) and unclear parachute shapes. For the MCGS sample in \autoref{fig:figure6}(d), the preservation of the overall shape was slightly worse than that of the TIES sample. The four rims were preserved with clear parachute shapes, and their importance order was completely maintained (a $>$ b $>$ c $>$ d).

%6.3.3
\subsubsection{Facebook1684 Graph Data Set}
The Facebook1684 graph data set is an online social network with 775 nodes and 14,006 edges. The original graph and six samples are shown in \autoref{fig:figure4}. This data set is an unbalanced graph in which two large communities include 705 nodes and two small communities/cliques contain only 70 nodes. This case focused on the preservation of ties between communities in an unbalanced graph. The SST (a), TIES (e), and RDN (f) samples maintained a few nodes in the two small communities and lost the connections between small and large communities. The FF (b) and RW (d) samples lost the two small communities and generated new chain-like rims at the margins of the largest communities. The MCGS sample (c) was the only one that effectively preserved the two small communities and the connections between small and large communities.

%sec7
\section{Discussion}
In this section, we discuss the limitations of this work and suggest directions for further work.
\subsection{Limitations}
{We mainly used scale-free graphs in this work, such as social, communication, and web networks. Whether our MCGS is applicable for other types of graphs, such as biological networks \cite{G29}, bipartite graphs \cite{G5}, and signed networks \cite{G6}, needs to be deliberated. Our definitions of the four minority structures could be different in other graph types.}

\par{The experimental study demonstrated that the existing algorithms had a low ability of preserving minority structures. However, two points should be noted. (1) Existing algorithms did not perform well just because they are not originally designed for minority structure preservation. They still had distinguished competences in diverse application scenarios \cite{G16}, such as RW in large graph estimation and RE in computational cost reduction. (2) Our MCGS is mainly suitable for graph analyses oriented to minority structures, such as graph visualization \cite{G19,addChen} and anomaly detection. Its ability for other scenarios remains unknown.}

\par{In the pilot user study, we ranked the eight SOI types only depending on the number of entries. In practice, the importance of SOI types should be considered. For example, BS type should rank first in network vulnerability analysis. Likewise, the definitions of minority structure types could be adjusted on demands \cite{G10,G38}. For example, HD-global structures are not needed to be subdivided into super pivots and huge stars in some cases. A tie was strictly defined as a single chain bridging two communities in this work, but communities may be connected by multiple chains.}

\par{In the evaluation, the experiment of majority structure preservation was not fully comprehensive. Some common metrics, such as clustering coefficient distribution and connected component size distribution \cite{A8,G32}, were not included. In the subjective assessment experiment, we presented six samples at a time to facilitate a convenient comparative perception, but this manner may cause differentiated ratings. An iterative manner is to show one sample at a time. {Moreover, we tested the proposed MCGS on large-scale graphs (see the supplementary material). The experimental results showed that the graph visualizations after sampling still presented severe visual clutters, even when the sampling rates were very low, which reflects that sampling in the data space may not be adequately suitable for visualizing large-scale graphs. It is worth to explore if sampling in the visual space can solve this challenge.}}

\subsection{Further work}
In the pilot user study, advanced techniques could be adopted in the future, such as using crowdsourcing approaches \cite{G13} to involve extensive participants and using eye tracking \cite{G14} to improve our manual SOI classification.

\par{In the experimental study, a ranking preservation indicator can be designed to evaluate the matching accuracy of the rankings of important minority structures before and after sampling, referring to the normalized discounted cumulative gain in the recommendation system community \cite{G3}. Sampling rates with a short interval and a wide range should be examined to find an appropriate sampling rate for minority structure preservation.}

\par{In the algorithm study, we plan to properly modify MCGS to extend its applied scope from undirected graphs to directed graphs or from scale-free graphs to other types of graphs \cite{G25}. We plan to add new types of minority structures and new objectives of loss function into the sampling process. Moreover, comprehensive methods for unbalanced graph identification and partition should be further studied. A layout that can present minority structures distinctly is worth further exploration \cite{G35,G37}.}

%删除G1,G30,G2
%Possible directions include matrix factorization methods \cite{G1,G30} and graph embedding methods \cite{G2}.

%motif analysis \cite{G11,G12}
%rare category discovery \cite{B55}
%Community margins and boundaries are sematic concepts \cite{G31}
%Fast Multipole Embedder \cite{G7} and Frutcherman Reingold \cite{G8}
%Force Atlas2 layout algorithm \cite{G39} 用到3.1去了

%sec8

\section{Conclusion}
{This work investigated the preservation of minority structures in graph sampling. We conducted a pilot user study and identified four representative types of minority structures. We conducted an experimental study and found that existing algorithms cannot effectively preserve the four types of minority structures. We designed a new graph sampling algorithm named MCGS that presented great performance of minority structure preservation in a series of experiments.} This work is the first investigation of minority structure preservation in graph sampling. We hope this work will be conducive to the research and application of graph analyses oriented to minority structures. We also expect that this work will inspire other researchers to further study minority structure classification, identification, sampling, and visualization.

\acknowledgments{
	We wish to thank Xin Sun from Adobe and Qiaomu Shi and Xinming Chen from Alibaba Cloud Intelligent Security Lab for the fruitful discussions. The work is supported in part by the National Key Research and Development Program of China (No. 2018YFB1700403) and the National Natural Science Foundation of China (No. 61872388, 61872389, 61872314, 61672538, and 61672308). MCGS at Github: \url{https://github.com/csuvis/MCGS}.}
%\newpage
%\bibliographystyle{abbrv}

%\bibliographystyle{unsrt}
\bibliographystyle{abbrv-doi}

%\bibliographystyle{abbrv-doi-narrow}
%\bibliographystyle{abbrv-doi-hyperref}
%\bibliographystyle{abbrv-doi-hyperref-narrow}
%%%%%%format
%%%%%%作者:<姓，名>的形势，姓在前名在后（按照谷歌学术）
%%%%%%题目/书名:首字母大写，":"后首字母大写，其余一律小写
%%%%%%期刊/会议名:实词一律大写
%%%%%%doi:能找到的都写上
%%%%%%会议是否加“proceedings of”:暂时没找到一个统一的标准，按照引文信息最完整的网站给的复制粘贴过来
%%%%%%出版社:@book,@inbook 加出版社，别的一律不加
\bibliography{myreference-doi}
\end{document}